\documentclass[journal,10pt,letterpaper]{IEEEtran}
\IEEEoverridecommandlockouts
\usepackage{graphicx,times,cite,amsmath, amssymb, amsthm, mathrsfs, epsfig, array, subfigure}
\usepackage{multirow}

\usepackage{cite}
\usepackage{lipsum} 
\usepackage{booktabs}

\usepackage{color}

\allowdisplaybreaks[2]
\newcommand{\figwidth}{\linewidth}

\usepackage[acronym]{glossaries} 
\newacronym{6G}{6G}{6th-generation}
\newacronym{5G}{5G}{5th-generation}
\newacronym{UAVs}{UAVs}{unmanned aerial vehicles}
\newacronym{HSRs}{HSRs}{high-speed railways}
\newacronym{LEO}{LEO}{low-earth orbit}
\newacronym{OTFS}{OTFS}{orthogonal time frequency space}
\newacronym{OFDM}{OFDM}{orthogonal frequency division multiplexing}
\newacronym{ICI}{ICI}{inter-carrier interference}
\newacronym{DD}{DD}{delay-Doppler}
\newacronym{TF}{TF}{time-frequency}
\newacronym{QoS}{QoS}{quality of service}
\newacronym{DF}{DF}{decode-and-forward}
\newacronym{AF}{AF}{amplify-and-forward}
\newacronym{NOMA}{NOMA}{non-orthogonal multiple access}
\newacronym{AWGN}{AWGN}{additive white Gaussian noise}
\newacronym{PSD}{PSD}{power spectral density}
\newacronym{i.i.d.}{i.i.d.}{independent and identically distributed}
\newacronym{SNR}{SNR}{signal-to-noise ratio}
\newacronym{ISFFT}{ISFFT}{inverse symplectic finite Fourier transform}
\newacronym{SFFT}{SFFT}{symplectic finite Fourier transform}
\newacronym{LDPC}{LDPC}{low-density parity-check}
\newacronym{MRC}{MRC}{maximum-ratio combining}
\begin{document}
\title{Outage Probability Analysis for OTFS \\with Finite Blocklength}
\markboth{}{}
\author{
\IEEEauthorblockN{Xin Zhang\IEEEauthorrefmark{1}, Wensheng Lin\IEEEauthorrefmark{1}, Lixin Li\IEEEauthorrefmark{1}, Zhu Han\IEEEauthorrefmark{2} and Tad Matsumoto\IEEEauthorrefmark{3}}\\
\IEEEauthorblockA{\IEEEauthorrefmark{1} School of Electronics and Information, Northwestern Polytechnical University, Xi'an, China \\
\IEEEauthorrefmark{2} University of Houston, Houston, USA \\
\IEEEauthorrefmark{3}IMT-Atlantic, France, and University of Oulu, Finland, Emeritus.  \\
Email: zxx\_@mail.nwpu.edu.cn; \{linwest, lilixin\}@nwpu.edu.cn; zhan2@uh.edu; tadeshi.matsumoto@oulu.fi} 

\thanks{Corresponding authors: Lixin Li, Wensheng Lin.
	
	This work was supported in part by National Natural Science Foundation of China under Grant 62101450, in part by the Young Elite Scientists Sponsorship Program by the China Association for Science and Technology under Grant 2022QNRC001, in part by Aeronautical Science Foundation of China under Grants 2022Z021053001 and 2023Z071053007, in part by the Open Fund of Intelligent Control Laboratory, 
	in part by NSF CNS-2107216, CNS-2128368, CMMI-2222810, ECCS-2302469, US Department of Transportation, Toyota and Amazon.}
}
\maketitle
\begin{abstract}
Orthogonal time frequency space (OTFS) modulation is widely acknowledged as a prospective waveform for future wireless communication networks.
To provide insights for the practical system design, this paper analyzes the outage probability of OTFS modulation with finite blocklength.
To begin with, we present the system model and formulate the analysis of outage probability for OTFS with finite blocklength as an equivalent problem of calculating the outage probability with finite blocklength over parallel additive white Gaussian noise (AWGN) channels.
Subsequently, we apply the equivalent noise approach to derive a lower bound on the outage probability of OTFS with finite blocklength under both average power allocation and water-filling power allocation strategies, respectively.
Finally, the lower bounds of the outage probability are determined using the Monte-Carlo method for the two power allocation strategies.
The impact of the number of resolvable paths and coding rates on the outage probability is analyzed, and the simulation results are compared with the theoretical lower bounds.
\end{abstract}
\begin{IEEEkeywords}
OTFS modulation, outage probability, finite blocklength, lower bound.
\end{IEEEkeywords}

\section{Introduction}
The \gls{6G} wireless communication networks are envisioned to provide ubiquitous connectivity for a diverse range of intelligent and highly mobile terminals, including autonomous vehicles, \gls{HSRs}, \gls{UAVs}, and \gls{LEO} satellites\cite{Wei2021a}.
Ensuring reliable and consistent performance in such high-mobility scenarios poses significant technical challenges for 6G networks.
However, high-mobility broadband signal transmission over wireless channels suffers from significant difficulties due to the doubly selective property of the channel, including frequency selectivity and time selectivity.
Mitigating time-selective fading caused by Doppler spread, which exceeds the channel's coherence time, is essential for ensuring robust and efficient communication in high-mobility scenarios.
Consequently, research has focused on developing novel modulation waveforms and architectures to support communication in high-mobility scenarios of 6G networks.

\Gls{OFDM} modulation has been widely adopted in \gls{5G} cellular networks and WiFi networks. 
However, OFDM exhibits significant limitations in high-mobility scenarios, as large Doppler shifts disrupt the orthogonality of subcarriers and introduce severe \gls{ICI}.
Recently, a novel two-dimensional modulation scheme, termed \gls{OTFS}, has been proposed as a prospective solution for high-mobility communication scenarios\cite{Hadani2017}.
Unlike traditional \gls{OFDM}, which modulates information symbols in the \gls{TF} domain, \gls{OTFS} maps the symbols onto the \gls{DD} domain. 
This transformation enables \gls{OTFS} to effectively convert doubly selective TF domain channels into quasistatic channels in the \gls{DD} domain. 
As a result, OTFS exhibits enhanced robustness against fading variations caused by high-mobility and provides improved performance in high-mobility scenarios with severe Doppler shifts\cite{Hadani2017a}.

The previously mentioned emerging technologies, such as autonomous vehicles, \gls{UAVs} and \gls{LEO} satellites, require not only seamless and reliable communication but also the ability to support services with varying \gls{QoS} requirements.
In such circumstances, guaranteeing reliable communication under fluctuating channel conditions and fulfilling diversity \gls{QoS} requirements poses a crucial challenge, which is typically evaluated by means of outage probability.
Several studies have explored outage probability analysis for OTFS systems and provided valuable insights into OTFS performance.

Chong et al.\cite{Chong2022a} analyze the outage performance of OTFS in both single-user and multi-user cases, deriving the lower bound of outage probability as a function of the achievable rates.
In reference \cite{Hu2022}, the outage probability of a multi-UAV-aided LEO satellite communication system employing the OTFS scheme is analyzed, under the UAVs' user independent motion model.
Closed-form expressions of the outage probability in OTFS cooperative communication systems, employing both \gls{DF} and \gls{AF} protocols, are presented in \cite{james}.
Nevertheless, existing research on the outage performance of OTFS typically operates under the ideal assumption of infinitely long blocklength, which does not reflect the physical constraints of finite blocklength in real-world communication systems.
Consequently, it is essential to investigate the outage probability of OTFS with finite blocklength to better align theoretical analysis with practical implementation.

Finite blocklength codes have been extensively studied as a foundation for short-packet communications, providing a critical approach to enabling low-latency services in communication systems.
In the landmark work \cite{finitecoing}, Polyanskiy et al. derive a closed-form expression for the decoding error probability under finite blocklength transmission over \gls{AWGN} channels.
Their results reveal that the decoding error probability for finite blocklength codes remains nonzero, even if the coding rate is below the Shannon capacity, and increases as the blocklength decreases.
Recently, analytical studies on finite blocklength performance have been extended to various scenarios, including cooperative communication systems\cite{Hu2018}, \gls{NOMA} schemes\cite{Ng2024} and green communications\cite{Liu2024}.
Furthermore, investigating the outage probability performance with finite blocklength provides a theoretical foundation for optimizing communication system latency and efficiency through blocklength allocation, power control and transmission rate selection\cite{Zhu2023}.

In light of the preceding discussion, we are motivated to investigate the outage probability of OTFS with finite blocklength.
The main contributions of this paper are summarized as follows:
(1) Modeling and analyzing the outage probability of OTFS with finite blocklength, providing the first investigation that accounts for blocklength constraints in practical OTFS communication systems;
(2) Deriving a lower bound on the outage probability of OTFS with finite blocklength through average power allocation by applying the equivalent noise approach;
(3) Incorporating the water-filling power allocation strategy into the analysis to obtain the lower bound of the outage probability under optimal power allocation across multiple parallel fading paths.
\section{System Model}\label{sec:Principle}

\subsection{Delay-Doppler Channels of OTFS}
\label{section:A}
The complex baseband channel impulse response $h(\tau,\nu)$ represents the DD domain channel's reaction to an impulse with a time delay $\tau$ and a Doppler shift $\nu$. 
Let $w(t)$ denote the \gls{AWGN} process, characterized by a one-sided power spectral density (PSD) of $N_0$.
Therefore, the signal $r(t)$, which is received at the receiver side with an input signal $s(t)$ over the DD domain channel, is expressed as follows:
\begin{align}
r(t)=\iint h(\tau,\nu)s(t-\tau)e^{j2\pi\nu(t-\tau)}d\nu d\tau+w(t).
\end{align}
In high user-terminal mobility scenarios, modelling the DD domain channel requires far fewer parameters than modelling the channel in the TF domain.
The reason can be attributed to the few number of reflectors, each associated with specific Doppler shifts and delays, within the transmission environment.
The sparsity and separability of the DD domain channel allow the channel response $h(\tau,\nu)$ to be expressed as:
\begin{align}
h(\tau,\nu)=\sum_{i\operatorname{=}1}^Lh_i\delta(\tau-\tau_i)\delta(\nu-\nu_i)\label{con:h_tau_nu}.
\end{align}
In (\ref{con:h_tau_nu}), $L$ denotes the number of independent resolvable paths, \emph{\text{$\delta(\cdot)$}} represents the Dirac delta function, and $h_i$, $\tau_i$, and $\nu_i$ refer to the complex fading coefficient, delay, and Doppler shift, respectively, associated with the \emph{i}-th path. 

Without loss of generality, let us consider an OTFS system with frame duration $NT$, where $N$ is the number of time slot, $T$ is the time elapsed in a slot. And each time slot has $M$ symblos.
Specifically, the delay and Doppler shift indices are denoted by $l_i$ and $k_i$, where
\begin{align}
\tau_i&=\frac{l_iT}{M},\:\nu_i=\frac{k_i}{NT}+\kappa_i.\label{con:kappa}
\end{align}
In most cases, the delay index $l_i$ lies within the interval $[0,l_{max}]$, where $l_{max}$ is generally less than or equal to $L_{CP}$. 
Here $L_{CP}$ denotes the length of the cyclic prefix, which is appended to the beginning of the sequence before the signal transmission.
Therefore, the maximum value of the channel delay is $\tau_{max}=\frac{l_{max}T}{M}$. 
Similarly, it is assumed that $k_i\in[-k_{max},k_{max}]$, which implies that the maximum value of the Doppler shift is $\nu_{max}=\frac{k_{max}}{NT}$.
In (\ref{con:kappa}), $-\frac{1}{2}<\kappa_i\leq\frac{1}{2}$ represents the fractional Doppler, which denotes the fractional offset relative to the nearest Doppler tap.
It is unnecessary to consider fractional delays, as the time axis resolution of $T/M$ is adequate to approximate the path delays to the nearest sampling points in typical wideband systems\cite{Raviteja2019a}.

Let us turn our attention to the time-varying channel fading coefficient.
For ease of expression, the channel fading coefficient vector is denoted as $\mathbf{h}=\left[h_1,h_2,\ldots,h_L\right]$, with a size of $1\times L$.
The elements in $\mathbf{h}$ are $h_i=a_i+jb_i$, where $a_i$ and $b_i$ are \gls{i.i.d.} Gaussian random variables with mean $\mu$ and variance $1/(2L)$.
Especially when $\mu=0$, $|h_i|$ follows the Rayleigh distribution.
Each channel fading coefficient corresponds to a delay and a Doppler shift, the latter of which is assumed to follow the classic Jakes spectrum with respect to $k_{max}$.
The channel coefficient, the delay and the Doppler spreads are mutually independent, and will be employed to construct channel models for simulations in the subsequent section.
\subsection{Outage Probability with Finite Blocklength}
Let 
Let $\mathbf{y}$ and $\mathbf{x}$ represent the DD domain transmitted and received symbol vectors, respectively. 
We now focus on the vector form representation of the input-output relationship in the DD domain,
\begin{align}
\mathbf{y}=\mathbf{H}_\mathrm{DD}\mathbf{x}+\mathbf{w},\label{con:yHDDxw}
\end{align}
where, $\mathbf{H}_\mathrm{DD}$ is the effective DD domain channel matrix 
\begin{align}
\mathbf{H}_{\mathrm{DD}}=\sum_{i=1}^Lh_i\left(\mathbf{F}_N\otimes\mathbf{I}_M\right)\mathbf{\Pi}^{l_i}\mathbf{\Delta}^{k_i+\kappa_i}\left(\mathbf{F}_N^\mathrm{H}\otimes\mathbf{I}_M\right),
\end{align}
with $\mathbf{F}_N$ being the discrete Fourier transform matrices of size $N\times N$, $\mathbf{\Pi}$ being a forward cyclic shift permutation matrix describing the delay effect, and $\boldsymbol{\Delta}=\operatorname{diag}\{\alpha^0,\alpha^1,\ldots,\alpha^{MN-1}\}$ being a diagonal matrix describing the Doppler effect with $\alpha=e^{\frac{j2\pi}{MN}}$\cite{Li2022}.
$\mathbf{w}$ in equation (\ref{con:yHDDxw}) represents the equivalent sample vector of AWGN in DD domain with one-sided PSD of $N_0$. 
In \cite[Section 4.4]{ddbook}, the OTFS input-output relationship in the time domain and the DD domain are derived, which proves that the noise in the DD domain remains \gls{AWGN}.

Given the blocklength and error probability, reference \cite{finitecoing} investigates the maximal achievable coding rate for Gaussian channels.
With blocklength $n$ and block error probability $\epsilon$, the finite blocklength coding rate is expressed by:
\begin{align}
R(n)=C-\sqrt{\frac{V}{n}}Q^{-1}(\epsilon)+\mathcal{O}\left(\frac{\mathrm{log}n}{n}\right),\label{con:R(n)}
\end{align}
where $Q^{-1}(\cdot)$ is the inverse Q-function, and the Q-function is typically defined as
\begin{align}
Q(x)=\int_{x}^{\infty}\frac{1}{\sqrt{2\pi}}e^{-t^{2}/2} dt. 
\end{align}
$C$ is the channel capacity and $V$ is the channel dispersion.
Specifically, for the one-dimentional AWGN channel, we have
\begin{align}
C=\frac{1}{2}\mathrm{log}_2(1+\gamma),\\
V=\frac{\gamma(2+\gamma)(\mathrm{log}e)^2}{2(1+\gamma)^2},
\end{align}
where, $\gamma$ denotes \gls{SNR}.

In this paper, once a block error occurs, it is treated as an outage event, i.e., block error rate equals outage probability.
Therefore, the outage probability with the finite blocklength can be calculated as:
\begin{align}
P_{out}=Q\left[\sqrt{\frac{n}{V}}\left(C-R(n)\right)\right].\label{con:Pout}
\end{align}
In particular, it should be noted that a strict approximation of the maximal achievable coding rate can already be achieved when blocklengths are greater than 100\cite{finitecoing}, and therefore the term $\mathcal{O}\left((\mathrm{log}n)/n\right)$ in (\ref{con:R(n)}) is neglected in this paper.
\section{Outage Performance Analysis}\label{sec:Coding}
The OTFS with finite channel coding system model is presented in Fig. 1. 
The original binary sequence of length $k$ is encoded at a rate $R_c$ to produce an codeword of length $n$.
Therefore, the channel coding rate can be represented as $R_c=k/n$.
Then, the codeword are mapped into one OTFS DD domain frame.
After applying \gls{ISFFT} and the Heisenberg transform, the frame is transmitted through the channel.
The channel consists of $L$ independent resolvable parallel AWGN paths.
At the receiver, the signal $r(t)$ is converted back into the DD domain through the application of the Wigner transform followed by the \gls{SFFT}.
The OTFS frame is subsequently reconstructed as a sequence of length $k$ with an outage probability $P_{out}$ through demapping and channel decoding.
Therefore, on the basis of the normalized achievable capacity of OTFS\cite{Chong2022a}, the outage probability of OTFS with finite blocklength is defined as 
\begin{align}
P_{out}^{theo}\triangleq\Pr\left\{\log_2\det\left(\!\mathbf{I}_{MN}+\frac{E_s}{N_0}\mathbf{H}_{\mathrm{DD}}^\mathrm{H}\mathbf{H}_{\mathrm{DD}}\!\right)\!<k\right\}.
\end{align}

\begin{figure}[!t]
\centering \includegraphics[width=1\figwidth]{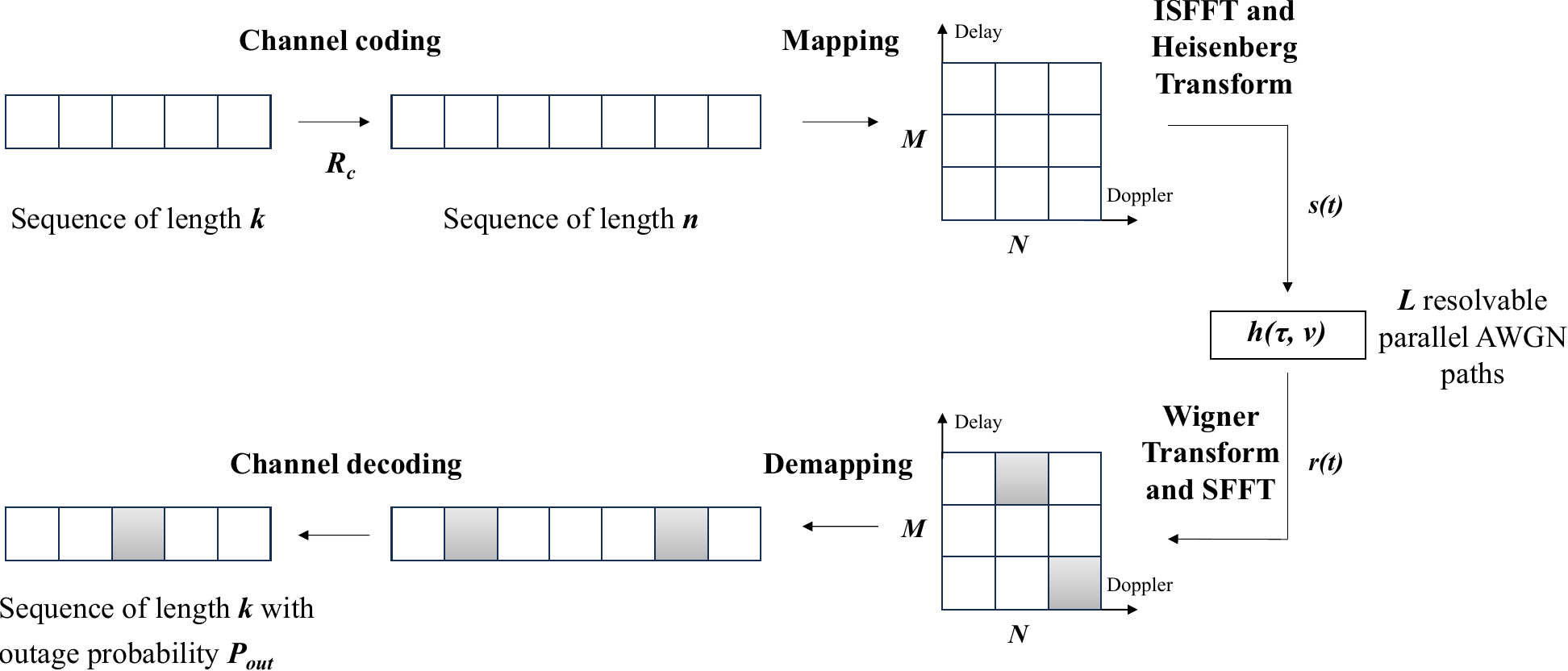}
\vspace{-2em}
\caption{The system model of OTFS with finite channel coding length.}
\label{fig:system_model}
\vspace{-1em}
\end{figure}

Fig. 2 illustrates the process of power allocation and SNR calculation for $L$-parallel AWGN paths in OTFS under finite blocklength conditions.
Each path is characterized by its channel coefficient $h_i$.
The transmit power allocated to the $i$-th path is denoted as $W_i$, and $W=\sum_{i=1}^LW_i$ denotes the total transmit power available for allocation to all $L$-parallel paths.
$\sigma_0^2$ represents the noise power at the receiver.
The remaining equations shown in Fig. 2 will be explained in the following paragraphs.

Reference \cite{parr_awgn} extends equation  (\ref{con:R(n)}) to the one-dimentional $L$-parallel AWGN channel.
The two-dimentional $L$-parallel AWGN channels can be equivalently represented as a superposition of one-dimentional $2L$-parallel AWGN channels.
Therefore, we can deduce the outage probability for OTFS with finite blocklength in two-dimentional $L$-parallel AWGN channels
\begin{align}
P_{out}=Q\left[\sqrt{\frac{n}{V_L(\alpha)}}\left(C_L(\alpha)-R_c\right)\right],\label{con:poutalp}
\end{align}
where
\begin{align}
C_L(\alpha)=\frac{1}{2}\sum_{i=1}^{2L}\mathrm{log}_2(1+\alpha_i),\label{con:CLa}\\
V_L(\alpha)=\frac{(\mathrm{log}e)^2}{2}\sum_{i=1}^{2L}\frac{\alpha_i(2+\alpha_i)}{(1+\alpha_i)^2}\label{con:VLa}.
\end{align}
Furthermore, $\alpha_i$ represents the \gls{SNR} of the $i$-th path, which is defined as $\alpha_i=\frac{W_i|h_i|^2}{\sigma_0^2}$.

Therefore, the problem of calculating the outage probability of OTFS with finite blocklength is reformulated as determining the outage probability for finite blocklength transmission over parallel AWGN channels. 
This reformulation assumes that all paths arrive at the receiver simultaneously, disregarding the adverse impact of different path delays on outage performance. 
Consequently, the derivation presented below represents the lower bound on the outage probability.

\begin{figure}[!t]
\centering \includegraphics[width=1\figwidth]{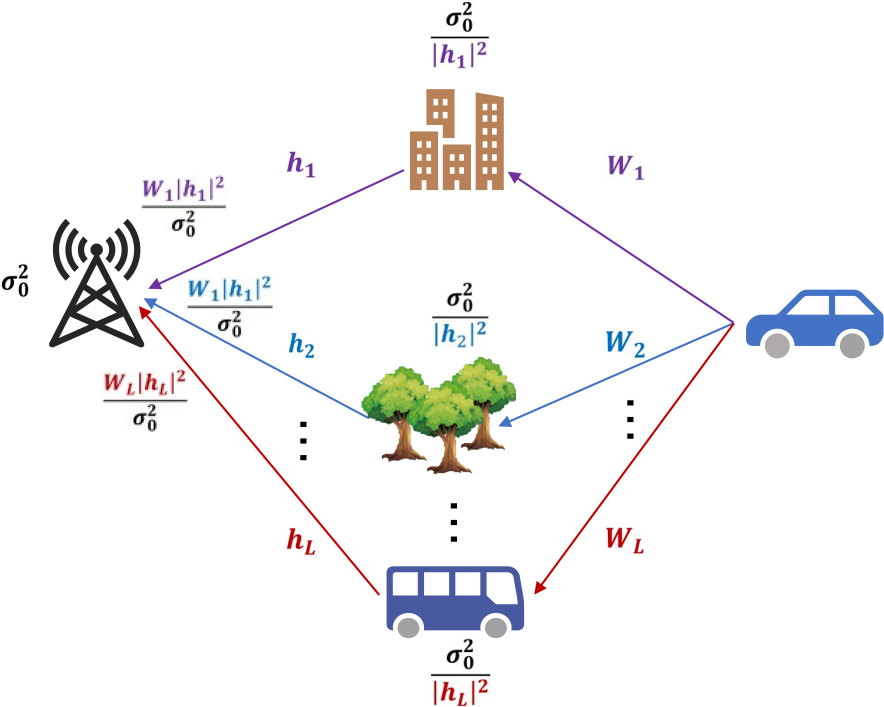}
\caption{Power allocation and SNR calculation for 
$L$-parallel AWGN paths in OTFS with finite blocklength.}
\label{fig:Eq_noise}
\vspace{-1em}
\end{figure}
Consider calculating the outage probability for OTFS with finite blocklength employing both the average power allocation and the water-filling power allocation.
For the average power allocation, it is simple to deduce that the SNR of the $i$-th path is $\alpha_i^{avg}=\frac{W|h_i|^2}{L\sigma_0^2}$.
By combinging (\ref{con:poutalp}) and $\alpha_i^{avg}$, the outage probability of OTFS with finite blocklength can be derived using the average power allocation.

For the water-filling power allocation, the equivalent noise will be calculated for each path.
To maxmize the overall \gls{SNR} across all paths, allocate higher power to paths with lower equivalent noise levels, and conversely, allocate lower power to paths with higher noise levels.
Given that the noise power at the receiver is identical for all paths, the equivalent noise is inversely proportional to the received signal power. Therefore, the equivalent noise for each path can be expressed as
\begin{align}
\varepsilon_i=\frac{\sigma_0^2}{|h_i|^2}.
\end{align}
The power allocation to each path can be determined based on the corresponding equivalent noise $\varepsilon_i$.
The power $P_i$ for each path is obtained using the conventional water-filling power allocation strategy, expressed as:
\begin{align}
P_i=
\begin{bmatrix}
\lambda-\varepsilon_i
\end{bmatrix}^+,
\end{align}
where $\lambda$ represents the water-filling level, which is the solution of the power constraint equation 
\begin{align}
\sum_{i=1}^LP_i=W.
\end{align}

Furthermore, the outage probability for OTFS with finite blocklength can
be derived using the water-filling power allocation strategy, as given by
\begin{align}
P_{out}^{wat}=Q\left[\sqrt{\frac{n}{V_L(\alpha)}}\left(C_L(\alpha)-R_c\right)\right], 
\end{align}
where $C_L(\alpha)$ and $V_L(\alpha)$ are defined in (\ref{con:CLa}) and (\ref{con:VLa}), respectively,  and $\alpha_i^{wat}=\frac{P_i|h_i|^2}{\sigma_0^2}$.
\section{Performance Evaluation}
In this section, the Monte-Carlo method\cite[Section 2.3]{Barbu2020} is employed to obtain the theoretical outage probability of OTFS with finite blocklength, as well as the outage probability lower bounds under average power allocation and water-filling power allocation.
A comparative analysis is conducted between the theoretical results and the derived lower bounds.
To maintain generality, we set $M=32$, $N=16$, ${\Delta f}=7.5\mathrm{KHz}$, $f_c=4\mathrm{GHz}$.
As discussed in Section \ref{section:A}, the channel coefficients are randomly generated following a complex Gaussian distribution with a mean of $\mu=0$ and a variance of $1/(2L)$. 
The maximum delay index is set to $8$, and the maximum Doppler index is set to $4$, corresponding to a maximum velocity of $500$ km/h. 

We first compare the lower bound of outage probability under average power allocation and water-filling power allocation for different numbers of resolvable paths, as illustrated in Fig. 3. 
For a given number of resolvable paths, it is evident that the optimal power allocation achieved by the water-filling allocation results in a lower outage probability compared to average power allocation. 
Furthermore, it is observed that in the low $E_s/N_0$ region, a higher number of resolvable paths exhibit higher outage probability than those with fewer paths. 
However, in the high $E_s/N_0$ region, the trend reverses, more resolvable paths achieving lower outage probability.
This phenomenon can be attributed to the fact that as the number of resolvable paths increases, the energy allocated to each individual path is relatively reduced, especially for average power allocation.
Moreover, a larger number of paths introduces additional interference.
However, in the high $E_s/N_0$ region, the diversity gain progressively becomes dominant, reducing the outage probability in cases with more paths.
\begin{figure}[!t]
\centering \includegraphics[width=1\figwidth]{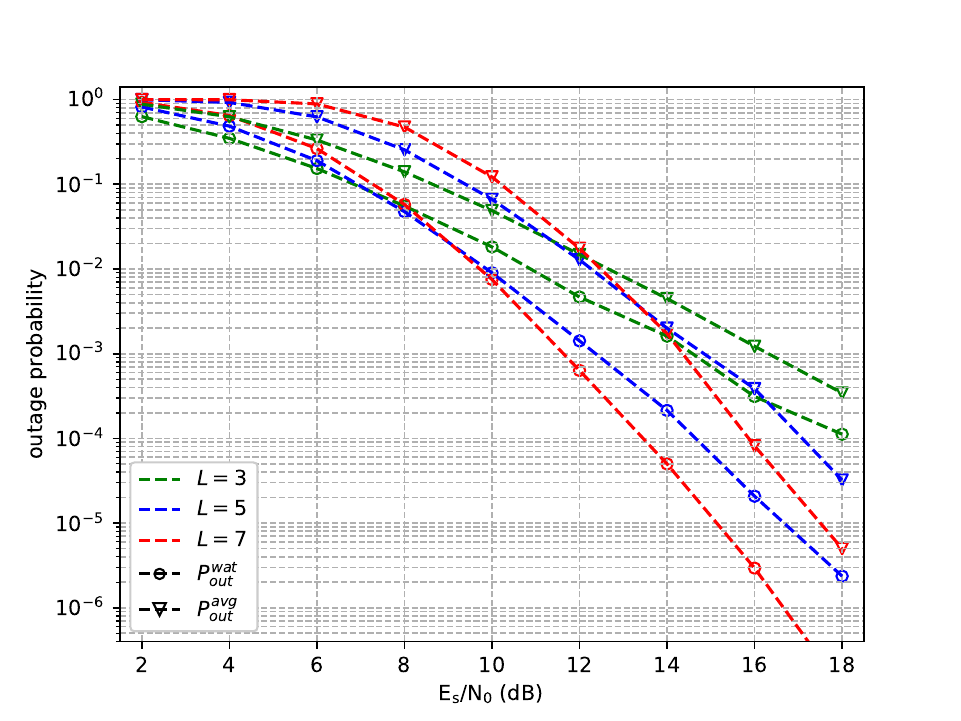}
\vspace{-2em}
\caption{Outage probability lower bounds for different number of resolvable paths under average and water-filling power allocation, where $R_c=0.8$.}
\vspace{-12pt}
\end{figure}

Fig. 4 illustrates the lower-bound curves of outage probability for average power allocation and water-filling power allocation at different coding rates with $L=5$.
It is obvious that lower coding rates result in smaller outage probability for the same $E_s/N_0$. 
It is also worth noting that in the high $E_s/N_0$ region, the slopes of the curves are essentially identical.

\begin{figure}[!t]
\centering \includegraphics[width=1\figwidth]{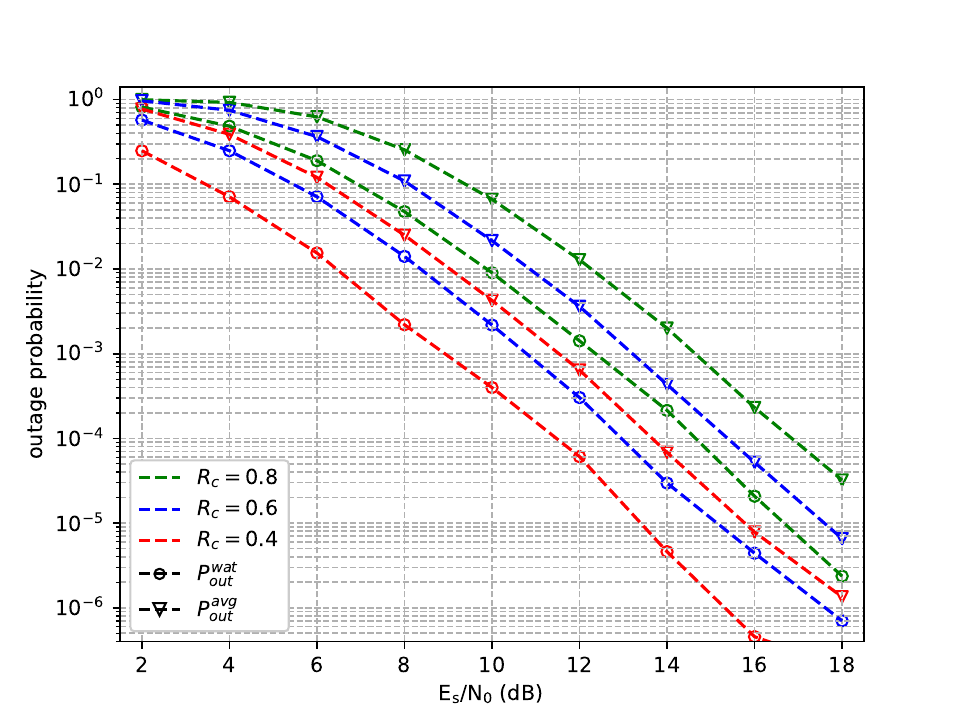}
\caption{Outage probability lower bounds for different coding rate under average and water-filling power allocation, where $P=5$.}
\vspace{-1em}
\end{figure}
Fig. 5 and Fig. 6 compare the theoretical results with the lower bound of outage probability under average power allocation for a coding rate of $0.8$, with $L=3$ and $L=5$, respectively.
Across the entire $E_s/N_0$ range, the outage probability lower bounds closely aligns with the theoretical results and exhibit a similar slope.
An increased number of independent resolvable paths reduces the gap between the theoretical results and the lower bounds.
It is evident that lower coding rates lead to improved outage probability performance. 
\begin{figure}[!t]
\centering \includegraphics[width=1\figwidth]{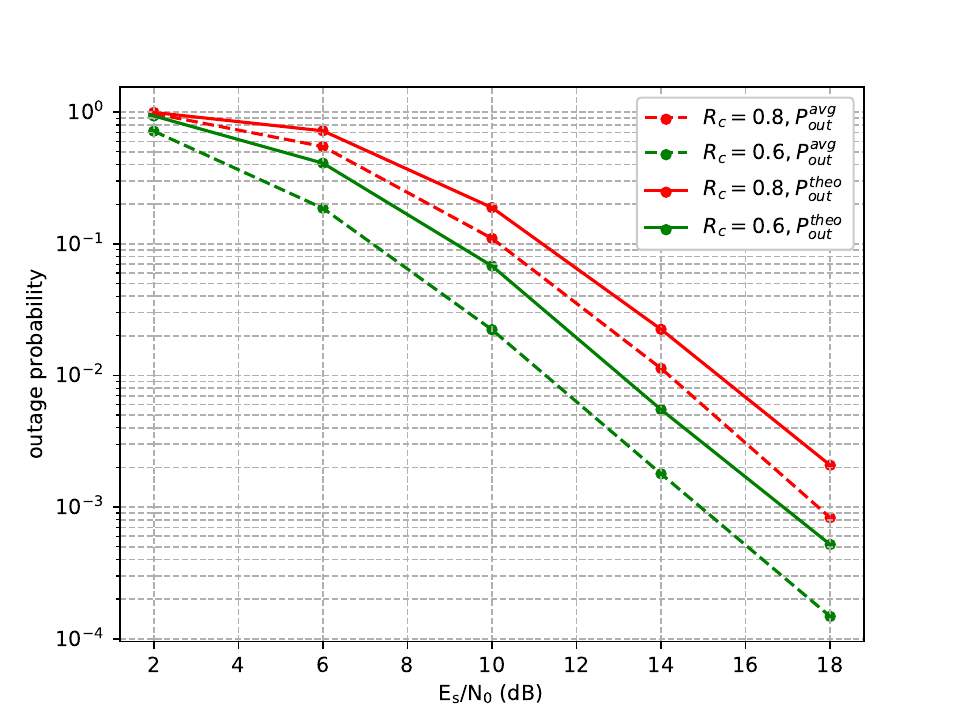}
\caption{Comparison of simulation and lower bound of outage probability with average power allocation, where $L=3$.}
\vspace{-15pt}
\end{figure}

\begin{figure}[!t]
\centering \includegraphics[width=1\figwidth]{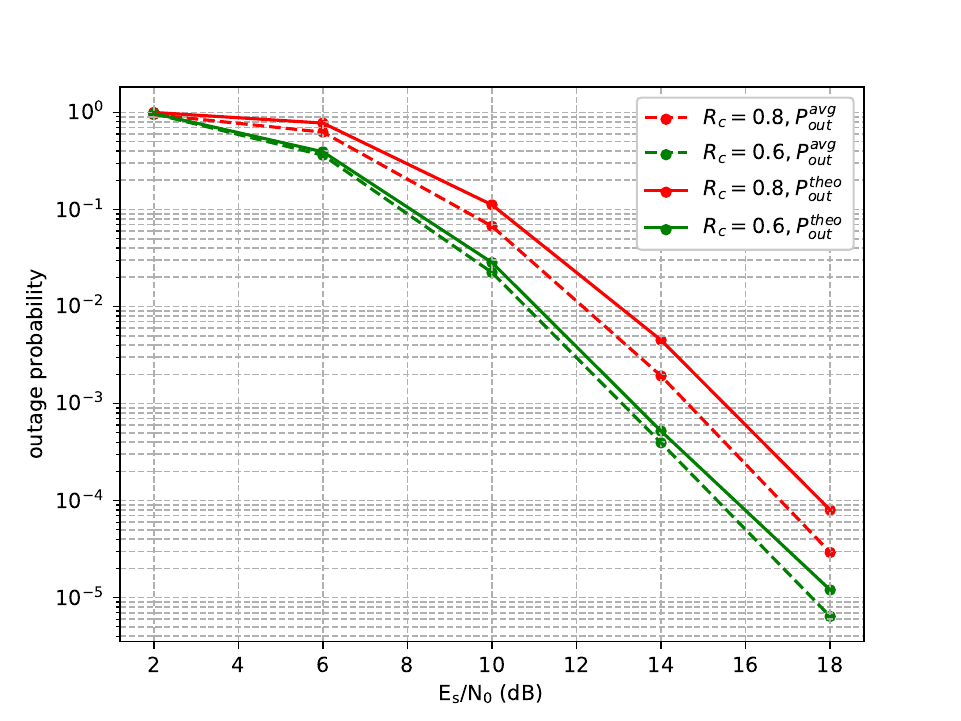}
\caption{Comparison of simulation and lower bound of outage probability with average power allocation, where $L=5$.}
\vspace{-10pt}
\label{fig:Eq_noise}
\end{figure}

\section{Conclusions}
\label{sec:Conclusion}
In this paper, we have conducted an analysis of the outage probability of OTFS modulation with finite blocklength.
We start from the system model of OTFS with finite channel coding rate.
Then, we derived theoretical lower bounds on the outage probability under both average and water-filling power allocation strategies, by formulating the problem as an equivalent analysis of finite blocklength over parallel AWGN channels. 
The performance evaluated through Monte-Carlo method reveal the impact of resolvable paths and coding rates on outage probability. 
Afterwards, the theoretical results are compared with the derived lower bounds.
Our analysis provides insights into the design and optimization of OTFS systems, specifically focusing on blocklength allocation, power control and transmission rate selection, to improve performance in high-mobility communication scenarios.
\bibliographystyle{IEEEtran}
\bibliography{reference}

\begin{thebibliography}{10}
\providecommand{\url}[1]{#1}
\csname url@samestyle\endcsname
\providecommand{\newblock}{\relax}
\providecommand{\bibinfo}[2]{#2}
\providecommand{\BIBentrySTDinterwordspacing}{\spaceskip=0pt\relax}
\providecommand{\BIBentryALTinterwordstretchfactor}{4}
\providecommand{\BIBentryALTinterwordspacing}{\spaceskip=\fontdimen2\font plus
\BIBentryALTinterwordstretchfactor\fontdimen3\font minus
  \fontdimen4\font\relax}
\providecommand{\BIBforeignlanguage}[2]{{%
\expandafter\ifx\csname l@#1\endcsname\relax
\typeout{** WARNING: IEEEtran.bst: No hyphenation pattern has been}%
\typeout{** loaded for the language `#1'. Using the pattern for}%
\typeout{** the default language instead.}%
\else
\language=\csname l@#1\endcsname
\fi
#2}}
\providecommand{\BIBdecl}{\relax}
\BIBdecl

\bibitem{Wei2021a}
Z.~Wei, W.~Yuan, S.~Li, J.~Yuan, G.~Bharatula, R.~Hadani, and L.~Hanzo,
  ``Orthogonal time-frequency space modulation: {A} promising next-generation
  waveform,'' \emph{IEEE Wireless Communications}, vol.~28, no.~4, pp.
  136--144, Aug. 2021.

\bibitem{Hadani2017}
R.~Hadani, S.~Rakib, M.~Tsatsanis, A.~Monk, A.~J. Goldsmith, A.~F. Molisch, and
  R.~Calderbank, ``Orthogonal time frequency space modulation,'' in \emph{IEEE
  Wireless Communications and Networking Conference (WCNC)}, San Francisco, CA,
  Mar. 2017, pp. 1--6.

\bibitem{Hadani2017a}
R.~Hadani, S.~Rakib, A.~F. Molisch, C.~Ibars, A.~Monk, M.~Tsatsanis,
  J.~Delfeld, A.~Goldsmith, and R.~Calderbank, ``Orthogonal time frequency
  space ({OTFS}) modulation for millimeter-wave communications systems,'' in
  \emph{IEEE MTT-S International Microwave Symposium (IMS)}, Honololu, HI, Jun.
  2017, pp. 681--683.

\bibitem{Chong2022a}
R.~Chong, S.~Li, W.~Yuan, and J.~Yuan, ``Outage analysis for {OTFS}-based
  single user and multi-user transmissions,'' in \emph{IEEE International
  Conference on Communications Workshops (ICC Workshops)}, Seoul, Republic of
  Korea, May 2022, pp. 746--751.

\bibitem{Hu2022}
J.~Hu, Y.~Jin, J.~Shi, X.~Liu, Z.~Dai, and Z.~Li, ``Reliability analysis of
  stochastic geometry-based multi-{UAV}-aided {LEO-Satcom} under {OTFS},'' in
  \emph{2022 International Symposium on Wireless Communication Systems
  (ISWCS)}, Hangzhou, China, Oct. 2022, pp. 1--6.

\bibitem{james}
A.~James and A.~Ananth, ``Outage analysis of orthogonal time frequency space in
  cooperative communication system,'' in \emph{2024 International Conference on
  Signal Processing and Communications (SPCOM)}, Bangalore, India, Jul. 2024,
  pp. 1--5.

\bibitem{finitecoing}
Y.~Polyanskiy, H.~V. Poor, and S.~Verdu, ``Channel coding rate in the finite
  blocklength regime,'' \emph{IEEE Transactions on Information Theory},
  vol.~56, no.~5, pp. 2307--2359, 2010.

\bibitem{Hu2018}
Y.~Hu, A.~Schmeink, and J.~Gross, ``Optimal scheduling of
  reliability-constrained relaying system under outdated {CSI} in the finite
  blocklength regime,'' \emph{IEEE Transactions on Vehicular Technology},
  vol.~67, no.~7, pp. 6146--6155, 2018.

\bibitem{Ng2024}
B.~K. Ng and C.-T. Lam, ``Characterization and optimization of coding
  performance in downlink {NOMA} with finite-alphabet inputs and finite
  blocklength,'' \emph{IEEE Transactions on Wireless Communications}, vol.~23,
  no.~4, pp. 2796--2811, 2024.

\bibitem{Liu2024}
Y.~Liu, J.~Lee, C.~Sun, Y.~Han, and W.~Chen, ``Energy efficient scheduling for
  short packet communications with finite blocklength coding,'' \emph{IEEE
  Transactions on Green Communications and Networking}, vol.~8, no.~4, pp.
  1645--1660, 2024.

\bibitem{Zhu2023}
Y.~Zhu, Y.~Hu, X.~Yuan, M.~C. Gursoy, H.~V. Poor, and A.~Schmeink, ``Joint
  convexity of error probability in blocklength and transmit power in the
  finite blocklength regime,'' \emph{IEEE Transactions on Wireless
  Communications}, vol.~22, no.~4, pp. 2409--2423, 2023.

\bibitem{Raviteja2019a}
P.~Raviteja, K.~T. Phan, and Y.~Hong, ``Embedded pilot-aided channel estimation
  for {OTFS} in delay–{Doppler} channels,'' \emph{IEEE Transactions on
  Vehicular Technology}, vol.~68, no.~5, pp. 4906--4917, May 2019.

\bibitem{Li2022}
S.~Li, W.~Yuan, Z.~Wei, and J.~Yuan, ``Cross domain iterative detection for
  orthogonal time frequency space modulation,'' \emph{IEEE Transactions on
  Wireless Communications}, vol.~21, no.~4, pp. 2227--2242, Apr. 2022.

\bibitem{ddbook}
Y.~Hong, T.~Thaj, and E.~Viterbo, \emph{Delay-{Doppler} Communications
  Principles and Applications}.\hskip 1em plus 0.5em minus 0.4em\relax
  Melbourne, VIC, Australia: Academic Press, 2022.

\bibitem{parr_awgn}
Y.~Polyanskiy, H.~V. Poor, and S.~Verdu, ``Dispersion of gaussian channels,''
  in \emph{2009 IEEE International Symposium on Information Theory}, Seoul,
  Republic of Korea, Jun. 2009, pp. 2204--2208.

\bibitem{Barbu2020}
A.~Barbu and S.-C. Zhu, \emph{{Monte} {Carlo} Methods}.\hskip 1em plus 0.5em
  minus 0.4em\relax Singapore: Springer, 2020.

\end{thebibliography}

\end{document}